\documentclass[pra,twocolumn,superscriptaddress,showpacs,preprintnumbers,amsmath,amssymb]{revtex4-1}

\usepackage{graphicx}
\usepackage{}
\usepackage{dcolumn}
\usepackage{bm}
\usepackage{ifpdf}
\usepackage{amsmath}

\begin{document}

\bibliographystyle{apsrev4-1} 


\title{Manipulating dipolar and spin-exchange interactions\\ in spin-1 Bose-Einstein condensates}
\author{Bo-Yuan Ning}
\affiliation{Department of Optical Science and Engineering, Fudan University, Shanghai 200433, China}%
\author{S. Yi}
\affiliation{Institute of Theoretical Physics, Chinese Academy of Sciences, Beijing 100190, China}%
\author{Jun Zhuang}
\affiliation{Department of Optical Science and Engineering, Fudan University, Shanghai 200433, China}%
\author{J. Q. You}
\affiliation{Department of Physics, Fudan University, Shanghai 200433, China}%
\author{Wenxian Zhang}%
\email{wenxianzhang@fudan.edu.cn}
\affiliation{Department of Optical Science and Engineering, Fudan University, Shanghai 200433, China}%
\date{\today}

\begin{abstract}
It remains a challenge to independently manipulate the magnetic dipolar and the spin-exchange interactions, which are entangled in many spin systems, particularly in spin-1 Bose-Einstein condensates. For this purpose, we put forward a sequence of rf pulses and the periodic dynamical decoupling sequence of optical Feshbach resonance pulses to control the dipolar and the spin-exchange interactions, respectively. Our analytic results and the numerical simulations demonstrate that either of the two interactions can be suppressed to make the other dominate the spin dynamics; furthermore, both of the interactions can be simultaneously suppressed to realize spinor-condensate-based magnetometers with a higher sensitivity. This manipulation method may find its wide applications in magnetic resonance and spintronics.
\end{abstract}


\maketitle

{\it Introduction---}
Long-range magnetic dipolar interaction and short-range spin-exchange interaction are fundamental in many spin systems, such as the electron and nuclear spin systems~\cite{Zhao2011}, the ferromagnetic ultrathin films~\cite{Won2003}, the ultracold atomic and molecular gases~\cite{Pethick2008, Ueda2010}, and so on. The competition between the two spin interactions diversifies not only the ground state but also the spin dynamics, especially in spinor ~\cite{Sengstock04,Chang04,Chang05,StamperKurn06nature,Ho1998,Ohmi1998,Zhang2005PRL} or dipolar Bose-Einstein condensates (BECs)~\cite{Pfau09,ZhangYB2005,Pfau07nature,KaiBongs07,Cherng09,Sau09,Bismut2010}, where the internal spin degrees of freedom are released in optical traps~\cite{Ketterle1998,Chapman01} and many physical parameters can be tuned precisely. However, the entanglement of the dipolar and spin-exchange interactions makes it difficult to understand the observed phenomena in many experiments. Such an interesting example is the spin texture or the spin domain in $^{87}$Rb spin-1 condensates, which might be ascribed to either one of these two interactions or both of them~\cite{StamperKurn08, Kawaguchi2010}.
Accordingly, it is highly desired to suppress one spin interaction and make the other dominant. Although efforts have been made to prohibit either the dipolar interaction~\cite{StamperKurn08} or the spin exchange interaction~\cite{Zhang2010, Ning2011}, it is yet difficult to clearly separate the individual effects.

\begin{figure}
\includegraphics[width=3.2in]{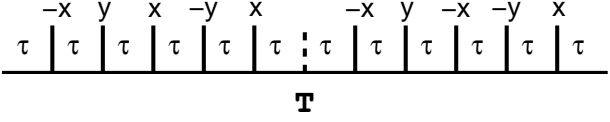}
\caption{Generalized WAHUHA pulse sequence of one period. The vertical lines mark the time of each applied rf pulse and the alphabets above the vertical lines represent the axis along which the spins are rotated $\pi/2$.}
\label{fig1}
\end{figure}

In this paper, we propose to {\it independently} suppress the dipolar and the spin-exchange interaction by a sequence of rf pulses ( Fig.~\ref{fig1}) and optical Feshbach periodic dynamical decoupling (PDD) sequences~\cite{Ning2011}, respectively. This scheme was verified by the analytic derivations, demonstrating that the manipulations of the relative strength between the two spin interactions enable experimentalists to unambiguously distinguish the specific role of the dipolar and the spin-exchange interactions. It is notable that the sequence of rf pulses is similar to the so called WAHUHA rf pulse sequence used to remove the secular dipolar interaction in nuclear spin systems~\cite{WAHUHA}, but our analytic results show that the WAHUHA sequence fails in spin-1 BECs while the pulse sequence shown in Fig.1, named as generalized WAHUHA (g-WAHUHA) rf pulse sequence in the following text, works well in any spin dipolar system. To verify the manipulation means under specific experimental conditions, we carried out numerical simulations with parameters of $^{87}$Rb spin-1 condensates and the results are in agreement with our analytic conclusions.

{\it The model---}
Within the mean field theory, coupled Gross-Pitaevskii equations for a spin-1 BEC in an arbitrary trap are\cite{MF,Kawaguchi06}
\begin{eqnarray}
\label{eq1}
\begin{split}
i\hbar\frac{\partial\psi_{\pm1}}{\partial t}=&\left[-\frac{\hbar^2}{2M}\nabla^2+V_{ext}+c_0N\pm c_2S_z\pm c_d\mathcal{D}_z\right]\psi_{\pm1} \\
&+\left[c_2S_{\mp}+c_d\mathcal{D}_{\mp}\right]\psi_0, \\
i\hbar\frac{\partial\psi_{0}}{\partial t}=&\left[-\frac{\hbar^2}{2M}\nabla^2+V_{ext}+c_0N\right]\psi_0 \\
& +\left[c_2S_++c_d\mathcal{D}_+\right]\psi_{+1} +\left[c_2S_-+c_d\mathcal{D}_-\right]\psi_{-1},
\end{split}
\end{eqnarray}
where $M$ is the atom mass, $\psi_{\alpha}$ ($\alpha=\pm1,0$) are the order parameters for the three components, $N=\sum_\alpha \psi^*_\alpha\psi_\alpha$ is the total number density, $\mathbf{S}= \sum_{\alpha,\beta} \psi^*_\alpha \mathbf{F}_{\alpha\beta} \psi_\beta$ is the spin density with $\mathbf{F}$ the atom spin matrix, and the dipole integral operator is $D_\eta(\mathbf{r})=\int d\mathbf{r}'\frac{1}{|\mathbf{r}-\mathbf{r}'|^3}[S_\eta(\mathbf{r}')-3e_\eta \mathbf{S}(\mathbf{r}')\cdot\mathbf{e}]$ ($\eta=x,y,z$) with $\mathbf{e}$ the unit dipolar vector; $S_{\pm}=(S_x\pm iS_y)/\sqrt{2}$ and $D_{\pm}=(D_x\pm iD_y)/\sqrt{2}$.
The spin-independent interaction and the spin-exchange interaction coefficients are, respectively, $c_0=4\pi\hbar^2(a_0+2a_2)/3M$ and $c_2=4\pi\hbar^2(a_2-a_0)/3M$ with $a_{0,2}$ denoting the $s$-wave scattering length in the two symmetric channels; the dipolar interaction coefficient is $c_d=\mu_0 g^2_F\mu^2_B/4\pi$ with $\mu_0$ the vacuum magnetic permeability, $\mu_B$ the Bohr magneton, and $g_F$ the Land\'{e} \textit{g}-factor.

We proceed by adopting the single mode approximation (SMA)~\cite{Law1998,Pu1999,SMA}, which separates the spinor condensate wave function into a spin-independent spatial part $\phi({\mathbf r})$ and the spin part $\vec{\xi}$, i.e., $\psi_\alpha(\mathbf r)=\phi(\mathbf r)\xi_\alpha$. By dropping the density dependent terms, which do not affect the spin dynamics, we obtain the Hamiltonian under SMA
\begin{equation}
\label{eq2}
\mathcal{H}_{SMA}=c_2'f^2+c_d'(3f_z^2-f^2)+3c_d''(f_y^2-f_x^2),
\end{equation}
where the isotropic spin-exchange term is $c_2'=(c_2/2)\int d\mathbf{r}\rho^2(\mathbf r)$ with $\rho(\mathbf r) = |\phi(\mathbf r)|^2$, the anisotropic dipolar term contains two parts, the cylindrically symmetric one $c_d'=(c_d/4)\int d\mathbf{r}d\mathbf{r'}\frac{1}{|\mathbf{r}-\mathbf{r'}|^3} \rho(\mathbf r)\rho(\mathbf{r'})(1-3\cos^2\theta_e)$ and the nonsymmetric one $c_d''=(c_d/4)\int d\mathbf{r}d\mathbf{r'} \frac{1}{|\mathbf{r}-\mathbf{r'}|^3} \rho(\mathbf{r})\rho(\mathbf{r'})\sin^2\theta_e e^{\pm i2\varphi_e}$ with $\theta_e$ and $\varphi_e$ being the polar and azimuthal angles of $(\mathbf{r}-\mathbf{r}')$. The amplitudes of $c_d'$ and $c_d''$ are approximately in the same order but are one order of magnitude smaller than that of $c_2$ for $^{87}$Rb condensates~\cite{Yi04}. The spins are $f_{x,y,z}=\langle\xi|F_{x,y,z}|\xi\rangle$ and $f^2=f_x^2+f_y^2+f_z^2$.
The spin dynamics under SMA becomes
\begin{eqnarray}
\label{eq3}
\begin{split}
i\hbar\frac{\partial}{\partial t}\xi_{\pm1}=&A_1[\xi_0^2\xi_{\mp1}^*+(n_0\pm n_{+1}\mp n_{-1})\xi_{\pm1}] \\
                                          \pm&A_2(n_{+1}-n_{-1})\xi_{\pm1} \\
                                          +&A_3[\xi_0^2(\xi_{+1}^*+\xi_{-1}^*)+n_0(\xi_{+1}+\xi_{-1})], \\
i\hbar\frac{\partial}{\partial t}\xi_{0}=&A_1[2\xi_0^*\xi_{+1}\xi_{-1}+\xi_0(n_{+1}+n_{-1})] \\
                                          +&A_3[\xi_0|\xi_{+1}+\xi_{-1}|^2+\xi_0^*(\xi_{+1}+\xi_{-1})^2], \\
\end{split}
\end{eqnarray}
where $A_1=2(c_2'-c_d'+3c_d'')$, $A_2=6(c_d'-c_d'')$, and $A_3=-6c_d''$. We have defined $\xi_\alpha=\sqrt{n_\alpha}\,e^{i\delta _\alpha}$ with $n_\alpha$ and $\delta_\alpha$ being the fractional population and the phase of the $\alpha$th component, respectively~\cite{Zhang05pra}.

{\it Independent suppression of the dipolar interaction---}
The Hamiltonian in Eq.(\ref{eq2}) reminds us two widely used methods, the magic angle and the WAHUHA sequence~\cite{WAHUHA}, which are capable of suppressing the dipolar interaction in nuclear spin systems. However, both of the two methods fail in a spin-1 BEC in that the nonsymmetric term with $c_d''$ in Eq.(\ref{eq2}) cannot be canceled~\cite{Personalcommu}.

We then generalize the WAHUHA sequence by adding extra $\pi/2$-rf pulses along $x$ direction (see Fig.~\ref{fig1}). Each pulse instantaneously rotate the spin $\pi/2$ along the designated direction. Due to its isotropic form, the spin-exchange term is not affected by any spin rotation, which provides us the opportunity to suppress the dipolar interaction independently. Below we show that this g-WAHUHA sequence indeed prohibits dipolar effects for an arbitrary shape condensate.
\begin{figure}
\includegraphics[width=3.5in,height=1.5in]{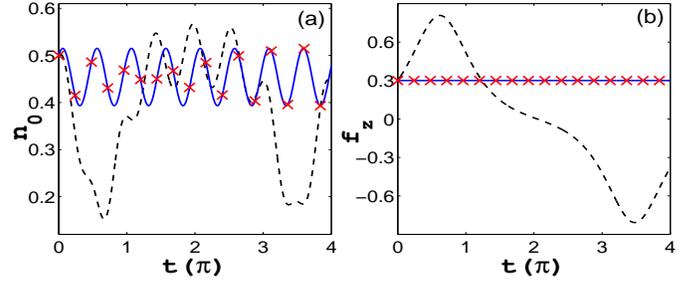}
\caption{(Color Online) Time evolutions of (a) $n_0$ and (b) $f_z$. The black dashed lines denote the free evolution with both spin-exchange and dipolar interactions, the blue solid lines denote the free evolution without dipolar interaction, and the red crosses denote the evolution under a sequence of 100 g-WAHUHA periods. The coincidence of the crosses and the solid lines clearly indicates that the dipolar interaction is suppressed by the applied g-WAHUHA sequence without affecting the spin-exchange effect. The initial values of ($n_{+1}, n_{0}, n_{-1}$) are (0.4, 0.5, 0.1) and the geometry of the trap is chosen as $c_d'=0.2c_2'$ and $c_d''=-0.1c_2'$ with $c_2'=-1$.
}
\label{fig2}
\end{figure}

We define the free evolution during interval $t\in [n \tau,(n+1)\tau]$ between two adjacent pulses as $U_0=\exp(-i\mathcal{H}_{SMA}\tau)$ and the rotation along the designated $\eta$ axis as  $U_\eta=\exp(-iF_\eta\,\frac{\pi}{2})$. The evolution for a complete g-WAHUHA period can be split into two halves, $U_T = U_{-}U_{+}$, where $U_{\pm}=U_0U_xU_0U_{-y}U_0U_{\pm x}U_0U_yU_0U_{-x}U_0$. In the toggling frame which is defined by the rf pulses~\cite{WAHUHA,Zhang08}, the first half-period evolution operator becomes $U_{+}=U_6U_5U_4U_3U_2U_1$ where $U_j=\exp(-iH_j \tau)$ ($j=1,2,\cdots, 6$) with
\begin{eqnarray}
\label{eq4}
H_1=c_2'f^2+c_d'(3f_z^2-f^2)+3c_d''(f_y^2-f_x^2), \nonumber \\
H_2=c_2'f^2+c_d'(3f_y^2-f^2)+3c_d''(f_z^2-f_x^2), \nonumber \\
H_3=c_2'f^2+c_d'(3f_x^2-f^2)+3c_d''(f_z^2-f_y^2), \nonumber \\
H_4=c_2'f^2+c_d'(3f_z^2-f^2)+3c_d''(f_x^2-f_y^2), \nonumber \\
H_5=c_2'f^2+c_d'(3f_y^2-f^2)+3c_d''(f_x^2-f_z^2), \nonumber \\
H_6=c_2'f^2+c_d'(3f_x^2-f^2)+3c_d''(f_y^2-f_z^2).
\end{eqnarray}

According to the average Hamiltonian theory~\cite{AH}, the evolution operator $U_{+}$ can be reexpressed as $U_{+}=\exp(-i\bar{H}\frac{T}{2})$ with $\bar{H}=\sum_{k=0}^\infty \bar{H}^{(k)}$. The term $\bar{H}^{(k)}$ is proportional to $\tau^k$ and can be obtained with Magnus expansion. In the limit of short $\tau$, the lowest order term $\bar{H}^{(0)}$, which is independent of $\tau$, becomes dominant
\begin{equation}
\label{eq5}
\begin{aligned}
\bar{H}^{(0)}=\frac{1}{6} \sum_{i=1}^6 H_i
             =c_2'f^2.
\end{aligned}
\end{equation}
Higher order terms are neglected since $\tau$ is small. In current experiments~\cite{StamperKurn08}, where the condensate density is about $10^{14}$cm$^{-3}$, $c_d'\sim c_d'' \sim 0.1 c_2' \sim 1$Hz, the pulse delay is $\tau \sim 10^{-3}$s, so that the higher order terms can be safely neglected, $\bar{H}^{(1)} \sim 10^{-4} \bar{H}^{(0)}$.

The average Hamiltonian given by Eq.(\ref{eq5}) clearly indicates that half of the g-WAHUHA sequence completely removes the dipolar interaction in the condensate to the lowest order $\tau$. Note that this result is universal, though the parameters $c_d'$ and $c_d''$ closely depend on the condensate shape. Finally, the toggling frame does not coincide with the laboratory frame because of the extra rotation term $U_xU_{-y}U_xU_yU_{-x}$, introduced by the applied five pulses. By simply adding five more pulses (see Fig.~\ref{fig1}), the second-half rotation term counteracts the first-half one. In this way, the toggling frame exactly coincides with the laboratory frame after a complete g-WAHUHA sequence.

To justify the above analysis of the g-WAHUHA method, we simulate the time evolution of a spin-1 system based on Eq.(\ref{eq3}) in three cases: (i) free evolution with both dipolar and spin-exchange interaction; (ii) free evolution with only the spin-exchange interaction; (iii) evolution with both spin interactions and g-WAHUHA sequence. If the results for case (ii) and (iii) are close, we safely conclude that the g-WAHUHA sequence effectively and independently suppresses the dipolar interaction.
We show in Fig.~\ref{fig2} $n_0$ and $f_z$, which are easily measured in experiments.
As can be clearly seen, the red crosses coincide with the blue solid lines, indicating the perfect agreement of our previous analysis and the numerical simulations, i.e., the g-WAHUHA sequence independently suppresses the dipolar interaction.
\begin{figure}
\includegraphics[width=3.5in,height=1.5in]{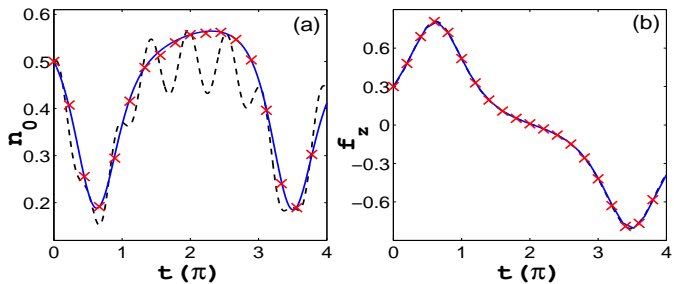}
\caption{(Color Online) Same as Fig.~\ref{fig2} except that the blue solid lines denote the free evolution without spin-exchange interaction and the red crosses denote the evolution under a sequence of 300 PDD pulses.The coincidence of the blue solid lines and the red crosses in both (a) and (b) indicates that the spin-exchange interaction are suppressed by the PDD sequence without affecting the dipolar effect. In (b), the black dashed line coincides with the blue solid one, manifesting the fact that the spin-exchange interaction does not affect the $f_z$ dynamics. }
\label{fig3}
\end{figure}

{\it Independent suppression of the spin-exchange interaction---}
We next consider the independent prohibition of the spin-exchange interaction in spin-1 BECs, with the presence of the dipolar interaction. Previous works demonstrate that PDD pulse sequences are able to suppress the spin dynamics induced solely by $c_2'$ term in Eq.(\ref{eq2}), by flipping the sign of $c_2$  through the optical Feshbach resonance technique~\cite{Hamley09,Zhang2010,Ning2011}. With the inclusion of dipolar terms, whether the PDD sequences perform the same as above is unknown. We show in the following within the average Hamiltonian theory that the optical PDD sequences indeed suppress the spin-exchange effects, independent of the inclusion of the dipolar interaction.

For a $2\tau$ PDD sequence, the evolution operator is $U_T = U_+U_-$ where $U_{\pm}=\exp(-iH_{\pm}\tau)$ with $H_{\pm} = \pm c_2' f^2 + c_d'(3f_z^2-f^2)+ 3c_d''(f_y^2-f_x^2)$. With the help of Magnus expansion, the average Hamiltonian to the lowest order is~\cite{AH}
\begin{equation}
\label{eq:pdd}
\bar{H}^{(0)} =  c_d'(3f_z^2-f^2)+ 3c_d''(f_y^2-f_x^2),
\end{equation}
which is free of $c_2'$ and $\tau$, and includes only the dipolar terms. Therefore, in the limit of small $\tau$, the optical PDD sequence removes the spin-exchange effects and leaves only the dipolar terms in a spin-1 condensate. Note that with a pure spin-exchange interaction, the PDD sequence can remove all the average Hamiltonian terms~\cite{Zhang2010}. Numerical results from Eq.~(\ref{eq3}) for $n_0$ and $f_z$ are shown in Fig.~\ref{fig3}. As one can see, the time evolution without spin exchange interaction (blue solid lines) is the same as that under the PDD sequence (red crosses), which confirms evidently the suppression of the spin-exchange interaction by the PDD sequence without affecting the dipolar effect.
\begin{figure}
\includegraphics[width=3.5in,height=1.5in]{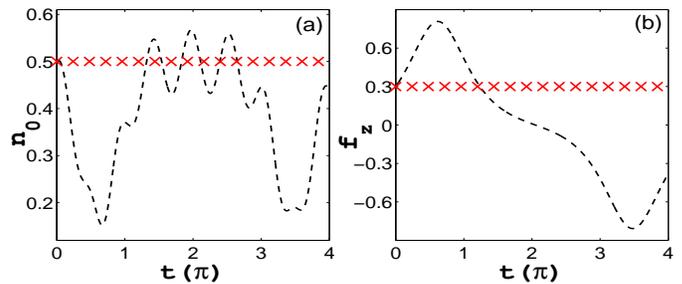}
\caption{(Color Online) Time evolution of (a) $n_0$ and (b)$ f_z$. The dashed lines denote the free evolution and the red crosses are under a sequence of 100 g-WAHUHA and PDD periods simultaneously. The freezing of the spin dynamics (crosses) indicates that both the dipolar and the spin-exchange interactions are suppressed. Parameters are the same as Fig.~\ref{fig2}.}
\label{fig4}
\end{figure}

\begin{figure}
\includegraphics[width=3.25in]{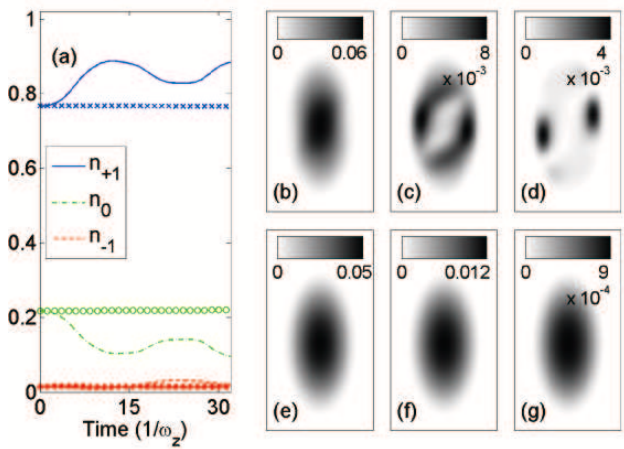}
\caption{(Color Online) Numerical simulations beyond SMA for a spin-1 BEC with spin-exchange and dipolar interactions. The system is either free or under the g-WAHUHA sequences. In (a): Time dependence of the fraction of each component, $n_{+1}$ (solid line), $n_0$ (dashed line), and $n_{-1}$ (dotted line) in free evolution; $n_{+1}$ (crosses), $n_0$ (circles), and $n_{-1}$ (asterisks) under the g-WAHUHA sequences [the pulse delay is $\tau=0.005(1/\omega_z)$]. The freely evolved density distributions in the $x-z$ plane at $t=30$ are shown in (b), (c), and (d) for $+1,0,-1$ components, respectively. The density distributions under g-WAHUHA sequences are shown correspondingly in (e), (f), and (g). The SMA is invalid for a freely evolved spin-1 condensate but it is kept if the g-WAHUHA sequences are applied, indicating the suppression of dipolar interactions. The initial state is a ground state of a $^{87}$Rb spin-1 condensate with pure spin-exchange interaction. The total atom number is $2\times10^5$, the initial fractions are $n_{+1}=0.767,n_0=0.218,n_{-1}=0.015$, and the aspect ratio is $\omega_x:\omega_y:\omega_z = 2:4:1$ with $\omega_z = (2\pi\times)~20$Hz. The dipolar coefficient $c_d$ is amplified five times to violate substantially the SMA during the free evolution.}
\label{fig5}
\end{figure}

{\it Simultaneous suppression of both dipolar and spin-exchange interactions---}
In some applications, for instance, a magnetometer based on spin-1 BECs~\cite{StamperKurn07}, it is desired to simultaneously suppress both spin interactions: the spin-exchange and the dipolar one. Since the g-WAHUHA sequence with rf pulses and the PDD sequence with optical ones are independent, it is straightforward to make a cooperative sequence which applies the rf pulses and the optical ones at the same time. For convenience, we set the period of the g-WAHUHA sequence to be the same as that of the PDD sequence. We present in Fig.~\ref{fig4} the frozen spin dynamics of the spin-1 condensates where the dipolar and the spin-exchange interactions are both suppressed.

{\it Beyond the single mode approximation---}
Our analysis are previously based on the SMA, which may not be valid in some experiments for spin-1 BECs, e.g., the spatial modes of each component are not the same in vortices, spin-domain or spin-texture experiments~\cite{StamperKurn08,Yi06,Kawaguchi06}. It has been shown that the optical PDD sequences can keep the initial SMA state from the spin domain induced by the spin-exchange interaction~\cite{Ning2011}. Similarly, we show in the following that the g-WAHUHA sequences also suppress the formation of spin spatial structure caused by the dipolar interaction.

The simulation starts from a SMA ground state with pure spin-exchange interaction, where the spin dynamics can be only introduced by the dipolar interactions. We turn on the dipolar interaction at $t=0$ and deliberately amplify its strength by five times to test the power of g-WAHUHA sequences. Two numerical simulations based on Eq.~(\ref{eq1}) are presented in Fig.~\ref{fig5} : In one simulation, we let the system evolve freely; in the other, we apply the g-WAHUHA sequences repeatedly. As shown clearly in Fig.~\ref{fig5}(a), the spin dynamics induced by the dipolar interaction is evidently suppressed and the fractions of each spin component are frozen, though the SMA is significantly violated during the free evolution [see Fig.~\ref{fig5}(b), (c), (d)].

{\it Conclusion---}
The proposed g-WAHUAHA sequence of rf pulses and the PDD seuqnces of optical Feshbach resonance pulses provide an experimentally feasible way to separate clearly the contribution of the dipolar and the spin-exchange interactions, which are usually entangled together in spin-1 BECs. Furthermore, by cooperatively suppressing both spin interactions, the precision of a magnetometer based on spin-1 Bose condensates can be significantly improved. Generally, the developed g-WAHUHA sequence for suppressing the dipolar interaction paves a new way to determine the atomic and molecular structures at zero or low magnetic fields in magnetic resonance imaging, nuclear magnetic resonance, chemistry, and biology. This sequence may also be utilized to extend the quantum coherence time of electron or nuclear spin systems in spintronics and quantum computing.

\begin{acknowledgements}
We thank L. You for helpful discussions. This work is supported by National Natural Science Foundation of China Grant Nos. 10904017, 11025421 (S.Yi) and 91121015(J. Q. You), NCET, Specialized Research Fund for the Doctoral Program of Higher Education of China under Grant No. 20090071120013, Shanghai Pujiang Program under Grant No. 10PJ1401300 and National Basic Research Program of China Grant No. 2009CB929300.
\end{acknowledgements}
%

\end{document}